\documentclass[12pt,preprint]{aastex}

\DeclareGraphicsExtensions{.eps}

\begin{document}

\title{On the structures in the afterglow peak emission of gamma ray bursts}

\author{Remo Ruffini\altaffilmark{1,2}, Carlo Luciano Bianco\altaffilmark{1,2}, Pascal Chardonnet\altaffilmark{1,3}, Federico Fraschetti\altaffilmark{1,4}, She-Sheng Xue\altaffilmark{1,2}}

\altaffiltext{1}{ICRA - International Center for Relativistic Astrophysics. E-mails: ruffini@icra.it, xue@icra.it}
\altaffiltext{2}{Physics Department, University of Rome ``La Sapienza", Piazzale Aldo Moro 5, I-00185 Rome, Italy. E-mail: bianco@icra.it}
\altaffiltext{3}{Universit\'e de Savoie, LAPTH - LAPP, BP 110, F­74941 Annecy-le-Vieux Cedex, France. E-mail: chardon@lapp.in2p3.fr}
\altaffiltext{4}{University of Trento, Via Sommarive 14, I-38050 Povo (Trento), Italy. E-mail: fraschetti@icra.it}

\begin{abstract}
Using GRB~991216 as a prototype, it is shown that the intensity substructures observed in what is generally called the ``prompt emission'' in gamma ray bursts (GRBs) do originate in the collision between the accelerated baryonic matter (ABM) pulse with inhomogeneities in the interstellar medium (ISM). The initial phase of such process occurs at a Lorentz factor $\gamma\sim 310$. The crossing of ISM inhomogeneities of sizes $\Delta R\sim 10^{15}\, {\rm cm}$ occurs in a detector arrival time interval of $\sim 0.4\, {\rm s}$ implying an apparent superluminal behavior of $\sim 10^5c$. The long lasting debate between the validity of the external shock model vs. the internal shock model for GRBs is solved in favor of the first.
\end{abstract}


To reproduce the observed light curve of GRB~991216, we have adopted, as initial conditions \citep{LongArticle} at $t=10^{-21}\, {\rm s}\sim 0\, {\rm s}$, a spherical shell of electron-positron-photon neutral plasma laying between the radii $r_0=6.03\times10^6\, {\rm cm}$ and $r_1=2.35\times10^8\, {\rm cm}$: the temperature of such a plasma is $2.2\, {\rm MeV}$, the total energy $E_{tot}=4.83\times 10^{53}\, {\rm erg}$ and the total number of pairs $N_{e^+e^-}=1.99\times 10^{58}$.

Such initial conditions follow from the EMBH theory we have recently developed based on energy extraction from a black hole endowed with electromagnetic structure (EMBH) \citep{rukyoto,prx98,rswx99,rswx00,brx00,lett1,lett2,lett3,LongArticle}, $r_0$ being the horizon radius, $r_1$ the dyadosphere radius and $E_{tot}$ coinciding with the dyadosphere energy $E_{dya}$. The above set of parameters is uniquely determined by the value of $E_{dya}$. The EMBH energy \citep{cr71} is carried away by a plasma of electron-positron pairs created by the vacuum polarization process \citep{dr75} occurring during the gravitational collapse leading to the EMBH \citep{crv02,rv02a}. Such an optically thick electron-positron plasma self propels itself outward reaching ultrarelativistic velocities \citep{rswx99}, interacts with the remnant of the progenitor star and by further expansion becomes optically thin \citep{rswx00}. The physical reason for such an extraordinary process of self-acceleration, achieving in a tenth of seconds in arrival time an increase in the Lorentz gamma factor from $\gamma=1$ to $\gamma\sim 300$, has been shown to be critically dependent on $E_{dya}$ and on the amount of baryonic matter engulfed by the plasma in its expansion \citep[see][]{rswx99,rswx00}. It is interesting that this process is extremely efficient even in the present case, regardless of the relatively slow random thermal motion of the $2.2\, {\rm MeV}$ $e^+e^-$ plasma \citep[see][]{LongArticle}. As the transparency condition is reached, a proper GRB (P-GRB) is emitted as well as an extremely relativistic shell of accelerated baryonic matter (ABM pulse). It is this ABM pulse which, interacting with the interstellar medium (ISM), gives origin to the afterglow \citep[see][]{lett2,LongArticle}.

One of the most novel results of the EMBH model has been the identification of what is generally called the ``prompt emission'' \citep[see e.g.][ and references therein]{p99} as an integral part of the afterglow: the extended afterglow peak emission (E-APE) \citep{lett1,lett2,LongArticle}. This result is clearly at variance with the models explaining the ``prompt emission" with ad-hoc mechanisms distinct from the afterglow process \citep[see e.g.][]{rm94,mr97b,kps97,rm98,mr00,rm00,kp00,m01}. The fact that the EMBH model, using GRB~991216 as a prototype, has allowed the computation of the temporal separation of the P-GRB and the E-APE to an accuracy of a few milliseconds and also to predict their relative intensities within a few percent can certainly be considered a major success of the model \citep[see][]{lett1,lett2,LongArticle}.

The aim of this letter is to report a further extension of the EMBH model in order to identify the physical processes giving origin to the intensity variability observed in the E-APE on time scales as short as a fraction of a second \citep{fm95}, which contrasts with the smoother emission in the last phases of the afterglow \citep[see e.g.][]{cfh01}.

In our former work on the EMBH model \citep{lett2,LongArticle}, we have assumed an homogeneous ISM with a density $n_{ism}=<n_{ism}>=1\, {\rm particle}/{\rm cm}^3$ and we have also assumed that during the collision of the ABM pulse with the ISM the ``fully radiative condition'' applies. These assumptions have led to the theoretical prediction of the power-law index of the afterglow slope $n=-1.6$ in excellent agreement with the observational data $n=-1.616 \pm 0.067$ \citep{ha00}. Our goal here is to show that the variability in the E-APE can indeed be traced back to inhomogeneities in the ISM. We again consider, like in the previous work, the case of an ABM pulse expanding with spherical symmetry (i.e. no beaming) and for simplicity we describe the ISM inhomogeneities as spherical shells concentric to the ABM pulse. Each shell has a selected density and a constant thickness $\delta R=1.0\times 10^{15}\, {\rm cm}$.

\begin{figure}
\plotone{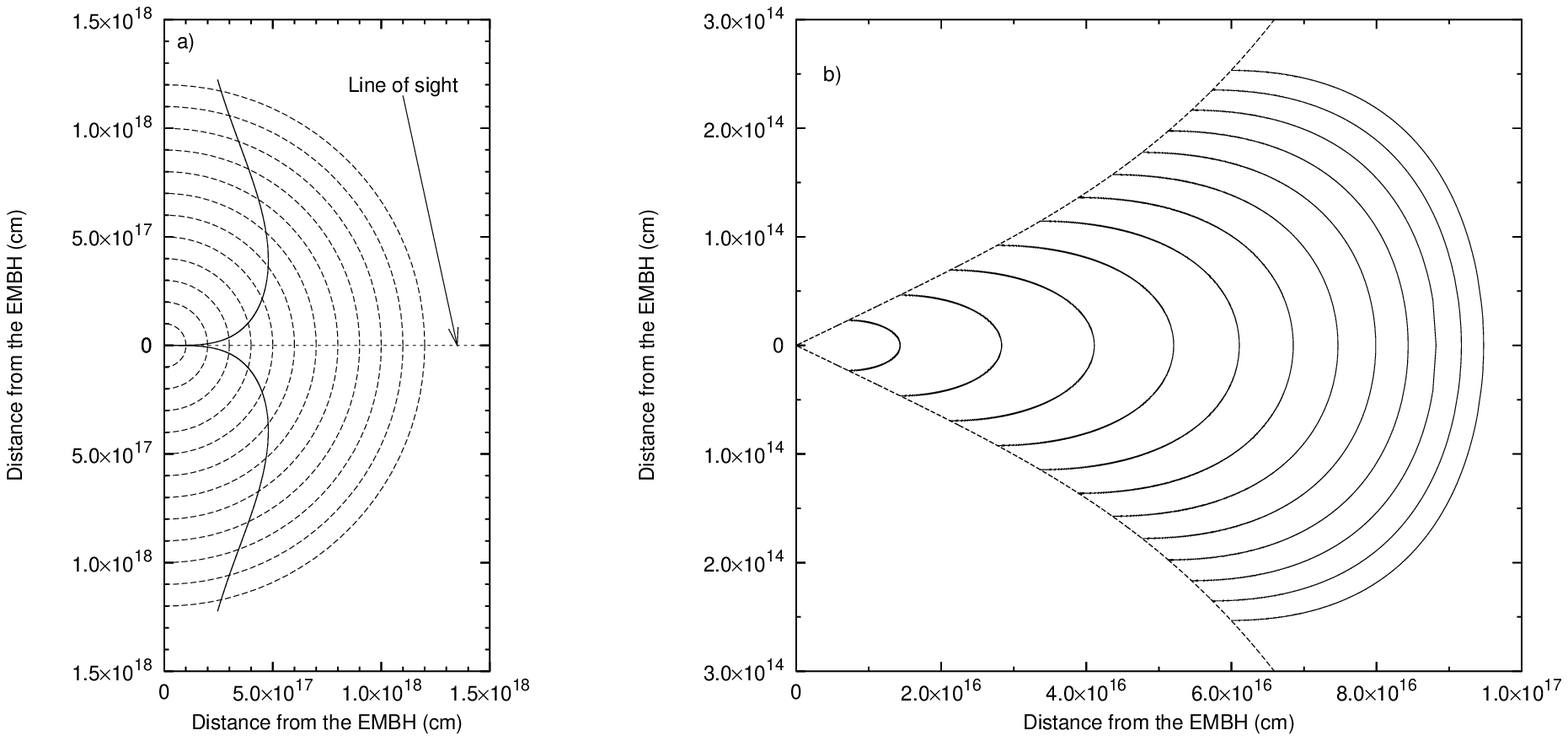}
\caption{{\bf a)} We represent the temporal evolution of the ABM pulse visible area. The dashed half-circles are the expanding ABM pulse at selected values of its radius corresponding to different laboratory times. The black curve marks the boundary of the visible region. The EMBH is located at position (0,0). The X (Y) axis is directed along (orthogonal to) the line of sight. In the earliest GRB phases the visible region is squeezed on the line of sight, while in the latest afterglow phases almost all the emitted photons reaches the observer. {\bf b)} In the same coordinate system used in a), we represent (solid lines) the {\em equitemporal surfaces} (EQTS) (see text). They corresponds to values of the arrival time ranging from $5\, s$ (the smallest surface on the left of the plot) to $60\, s$ (the largest one on the right) in steps of $5\, s$. The dashed lines are the boundaries of the ABM pulse visible area. Note the different scale on the two axis, indicating the very high EQTS ``effective eccentricity". The arrival time interval has been chosen to encompass the E-APE emission, occurring between $\sim 15\, {\rm s}$ and $\sim 40\, {\rm s}$ (see Figs.~\ref{grb991216}--\ref{substr_peak} and Tab.~\ref{tab1}).\label{opening_ETSNCF}}
\end{figure}

We recall now the relation between the relativistic beaming angle and the arrival time of the emitted photon on the detector. The visible part of the ABM pulse spherical surface is constrained by:
\begin{equation}
\cos\vartheta\ge {v\over c},
\label{cos}
\end{equation}
where $\vartheta$ is the angle in the laboratory frame between the radial direction of each point on the ABM pulse surface and the line of sight and $v$ is the expansion speed \citep{LongArticle2}. This follows from the requirement that in the comoving frame the component of the photon momentum along the radial expansion velocity direction be positive, in order to escape. There exists then a maximum allowed $\vartheta$ value $\vartheta_{max}$ defined by $\cos\vartheta_{max}=\left(v/c\right)$ (see Fig.~\ref{opening_ETSNCF}a).

\begin{figure}
\plotone{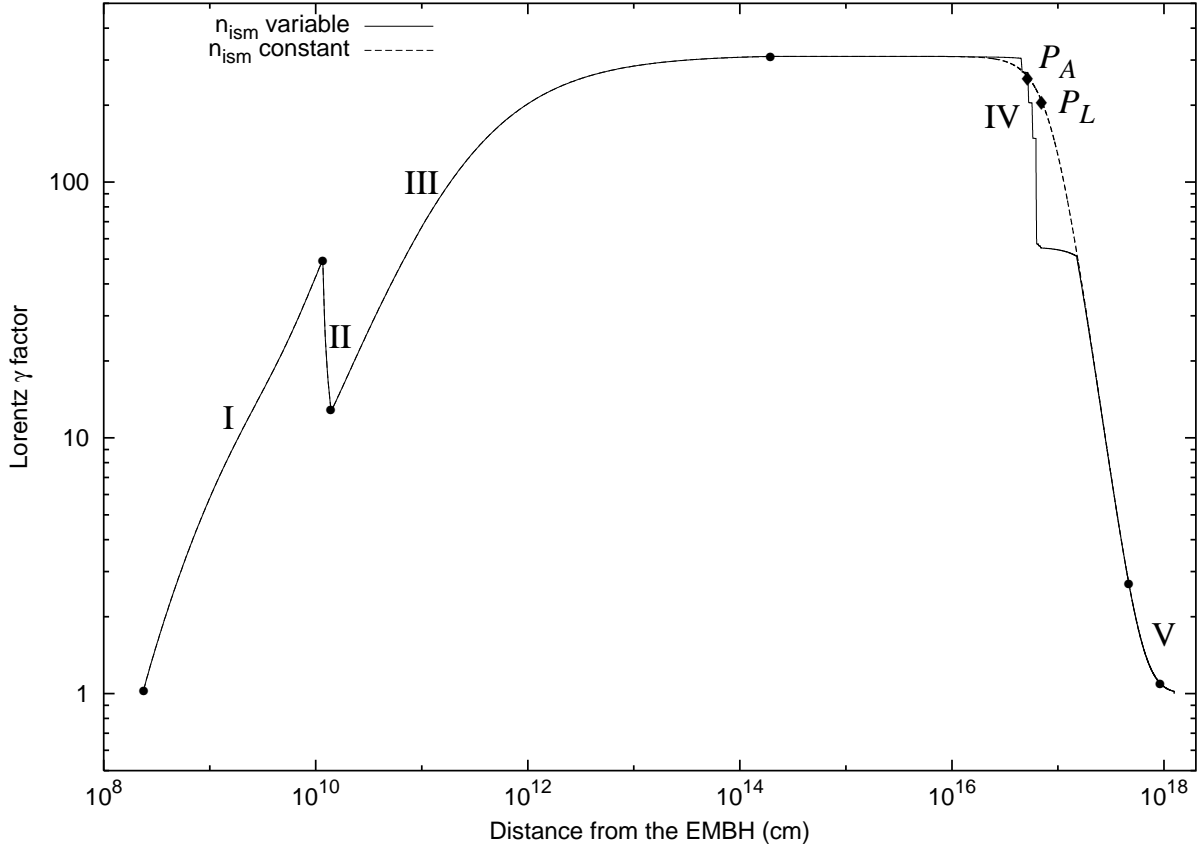}
\caption{The theoretically computed gamma factors of the expanding pulse are given as a function of its distance from the EMBH. The dashed line corresponds to $n_{ism}=1$ while the solid line corresponds to the density profile given in Fig.~\ref{maschera}. The roman numerals corresponds to the different eras of the EMBH theory \citep[see][]{LongArticle}. Near the E-APE (namely around $P_A$ and $P_L$) the two curves differs markedly, due to the impact on the high density ISM regions which brakes the ABM pulse more efficiently. When the ABM pulse overcomes the ISM cloud the two curves coincide again, since the average density of the cloud is $\sim 1\, {\rm particle}/{\rm cm}^3$. \label{gamma_sub}}
\end{figure}

Due to the high value of the Lorentz $\gamma$ factor $\left(\sim 310\right)$ for the bulk motion of the ABM pulse, the spherical waves emitted from its external surface do appear extremely distorted to a distant observer. To show this we need to express the photon arrival time at the detector $t_a^d$ of a function of its emission time $t$ and angle $\vartheta$. We set $t=0$ when the plasma starts to expand, so that $r\left( 0 \right)=r_{\rm ds}$. We then have \citep[see][]{LongArticle2}:
\begin{equation}
t_a^d  =\left(1+z\right)\left(t - \frac{{\int_0^t {v\left( {t'} \right)dt'}  + r_{\rm ds} }}{c}\cos \vartheta  + \frac{{r_{\rm ds} }}{c}\right)\, ,
\label{tad_fin}
\end{equation}
where $z$ is the redshift of the source. Then, in order to compute the arrival time of the emitted radiation, we must know {\em all} the previous values of the source velocity starting from $t=0$. The great advantage of the EMBH model is that for the first time we have been able to obtain the precise values of the gamma Lorentz factor as a function of the radial coordinate or equivalently of the laboratory time (see Fig.~\ref{gamma_sub}). This allows us, for the first time, to evaluate Eq.(\ref{tad_fin}) and correspondingly determine the surfaces that emits the photons detected at a fixed arrival time $t_a^d$, which we will call in the following ``equitemporal surfaces'' (EQTS). The profiles of such surfaces are reported in Fig.~\ref{opening_ETSNCF}b. We emphasize once again the direct connection between the evaluation of the EQTS and the {\em entire} past history of the source.

\begin{figure}
\plotone{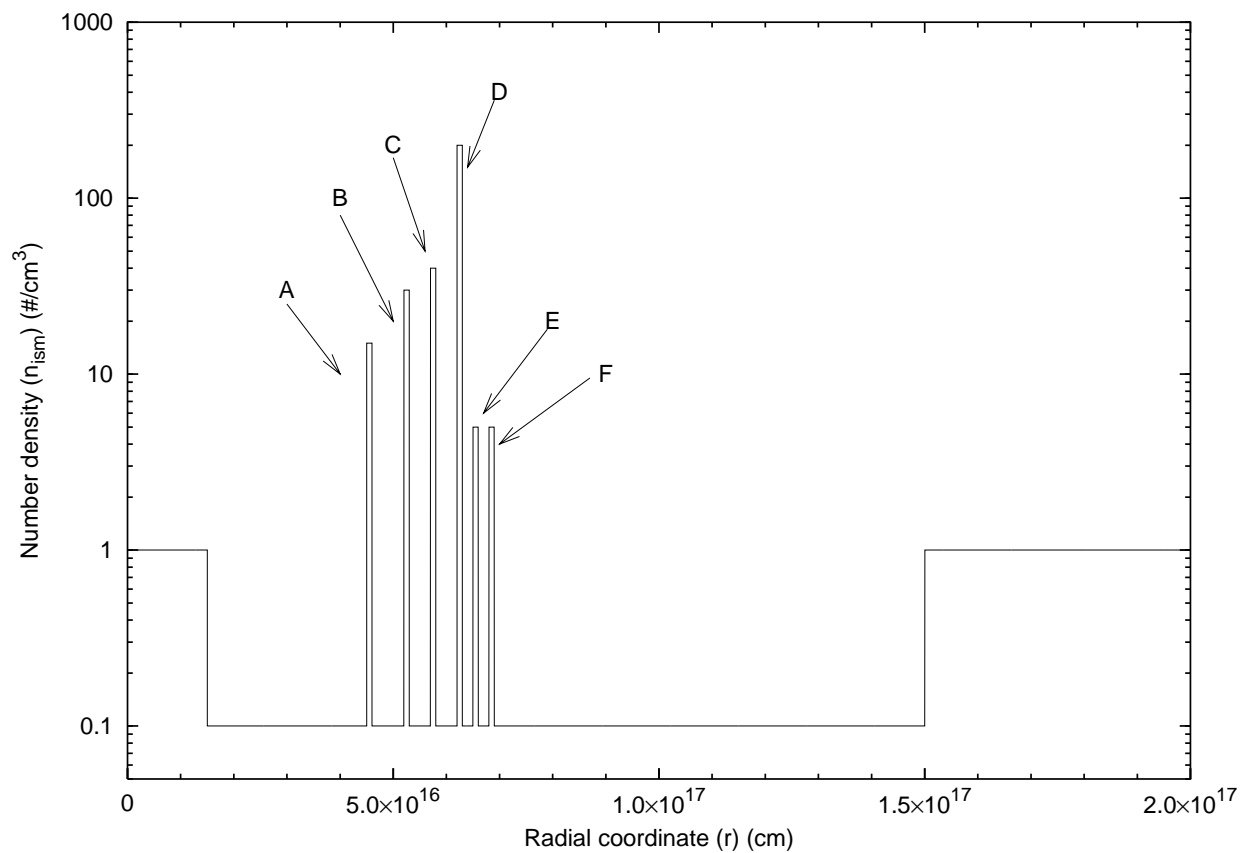}
\caption{The density profile (``mask") of an ISM cloud used to reproduce the GRB~991216 temporal structure. As before, the radial coordinate is measured from the black hole. In this cloud we have six ``spikes" with overdensity separated by low density regions. Each spike has the same spatial extension of $10^{15}\, {\rm cm}$. The cloud average density is $<n_{ism}>=1\, {\rm particle}/{\rm cm}^3$.\label{maschera}}
\end{figure}

We have created an ISM inhomogeneity ``mask" (see Fig.~\ref{maschera} and Tab.~\ref{tab1}) with the main criteria that the density inhomogeneities and their spatial distribution still fullfill $<n_{ism}>=1\, {\rm particle}/{\rm cm}^3$.

The source luminosity in a detector arrival time $t_a^d$ and per unit solid angle $d\Omega$ is given by \citep[details in][]{LongArticle2}:

\begin{equation}
\frac{dE_\gamma}{dt_a^d d \Omega } = \int_{EQTS} \frac{\Delta \varepsilon}{4 \pi} \; v \; \cos \vartheta \; \Lambda^{-4} \; \frac{dt}{dt_a^d} d \Sigma\, .
\label{fluxarr}
\end{equation}
where $\Delta \varepsilon$ is the energy density released in the interaction of the ABM pulse with the ISM inhomogeneities measured in the comoving frame, $\Lambda=\gamma(1-(v/c)\cos\vartheta)$ is the Doppler factor and $d\Sigma$ is the surface element of the EQTS at detector arrival time $t_a^d$ on which the integration is performed. In the present case the Doppler factor $\Lambda^{-4}$ in Eq.(\ref{fluxarr}) enhances the apparent luminosity of the burst, as compared to the intrinsic luminosity, by a factor which at the E-APE is in the range between $10^{10}$ and $10^{12}$!

\begin{figure}
\plotone{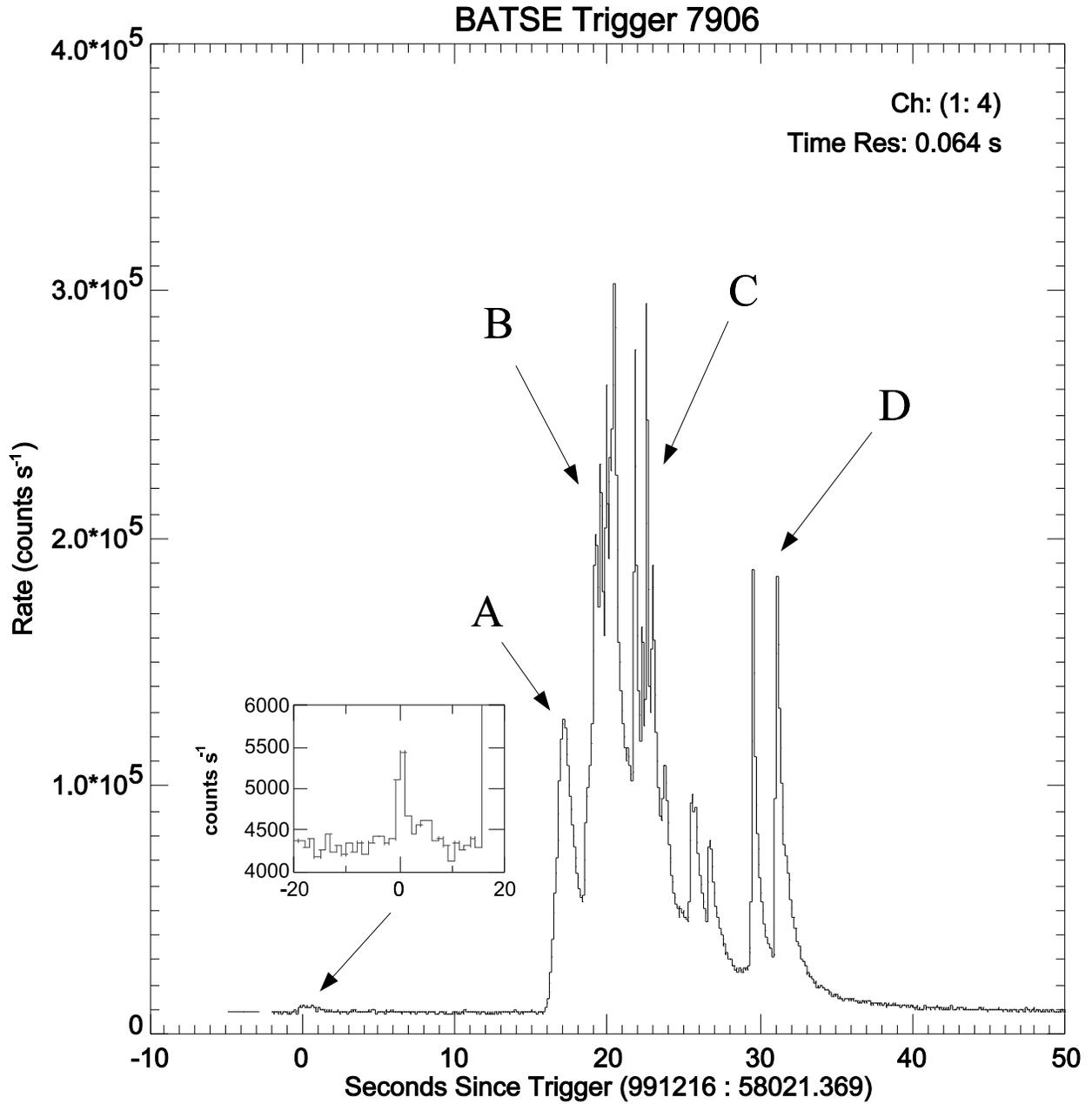}
\caption{The BATSE data on the E-APE of GRB~991216 \citep[source:][]{grblc99} together with an enlargement of the P-GRB data \citep[source:][]{brbr99}. For convenience each E-APE peak has been labeled by a different uppercase Latin letter. \label{grb991216}}
\end{figure}

\begin{figure}
\plotone{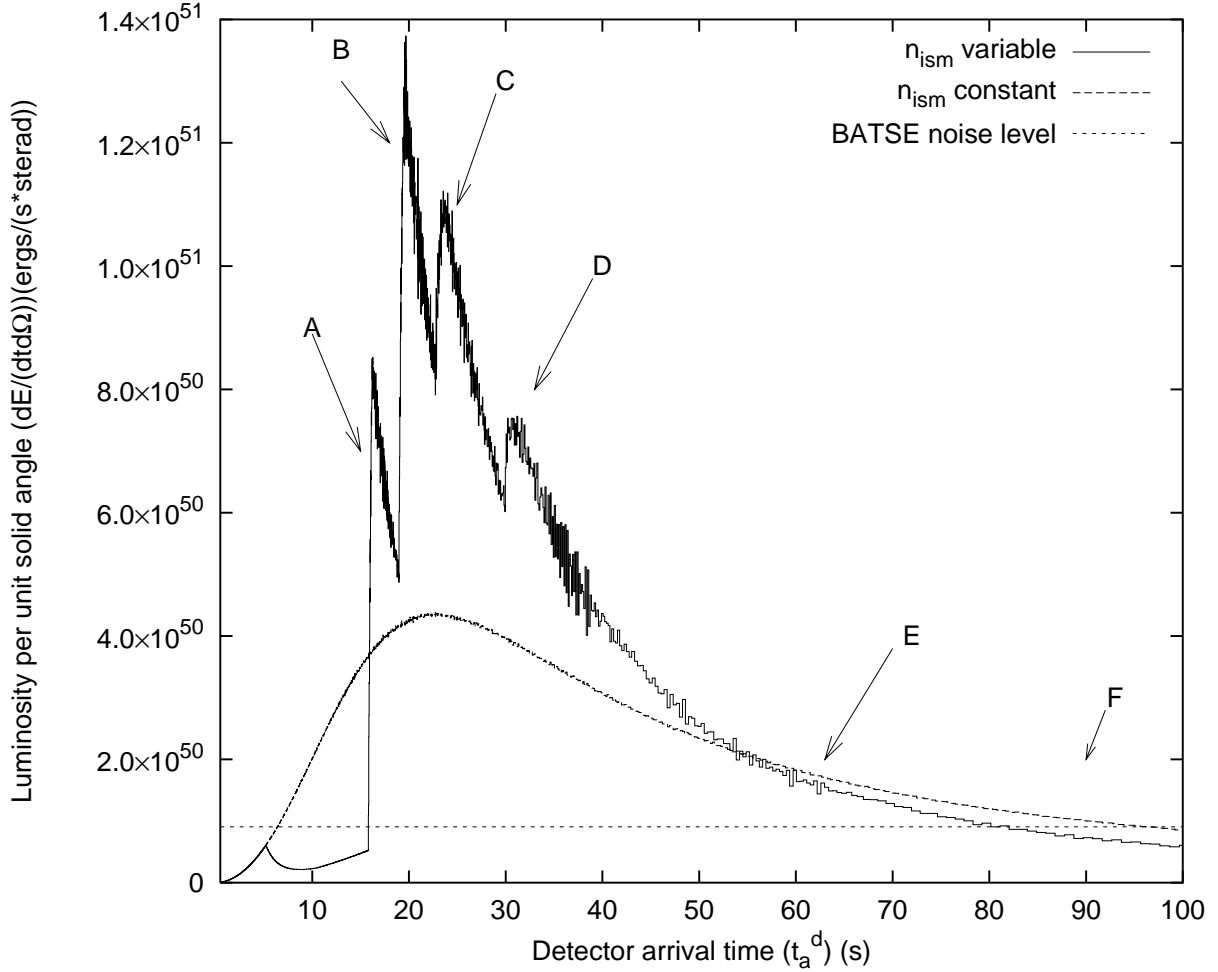}
\caption{The source luminosity connected to the mask in Fig.~\ref{maschera} is given as a function of the detector arrival time (solid ``spiky" line) with the corresponding curve for the case of constant $n_{ism}=1\, {\rm particle}/{\rm cm}^3$ (dashed smooth line) and the BATSE noise level (dotted horizontal line). The ``noise" observed in the theoretical curves is due to the discretization process adopted, described in \citet{LongArticle2}, for the description of the angular spreading of the scattered radiation. For each fixed value of the laboratory time we have summed $500$ different contributions from different angles. The integration of the equation of motion of this system is performed in $22,314,500$ contributions to be considered. An increase in the number of steps and in the precision of the numerical computation would lead to a smoother curve. \label{substr_peak}}
\end{figure}

The results are given in Fig.~\ref{substr_peak}. We obtain, in perfect agreement with the observations (see Fig.~\ref{grb991216}):
\begin{enumerate}
\item the theoretically computed intensity of the A, B, C peaks as a function of the ISM inhomogneities;
\item the fast rise and exponential decay shape for each peak;
\item a continuous and smooth emission between the peaks.
\end{enumerate}

Interestingly, the signals from shells E and F, which have a density inhomogeneity comparable to A, are undetectable. The reason is due to a variety of relativistic effects and partly to the spreading in the arrival time, which for A, corresponding to $\gamma=303.8$ is $0.4 s$ while for E (F) corresponding to $\gamma=57.23$ $(56.24)$ is of $10.2\, {\rm s}$ $(10.6\, {\rm s})$ (see Tab.~\ref{tab1} and \citet{LongArticle2}).

In the case of D, the agreement with the arrival time is reached, but we do not obtain the double peaked structure. The ABM pulse visible area diameter at the moment of interaction with the D shell is $\sim 1.0\times 10^{15}\, {\rm cm}$, equal to the extension of the ISM shell \citep[see Tab.~\ref{tab1} and][]{LongArticle2}. Under these conditions, the concentric shell approximation does not hold anymore: the disagreement with the observations simply makes manifest the need for a more detailed description of the three dimensional nature of the ISM cloud.

The physical reasons for these results can be simply summarized: we can distinguish two different regimes corresponding in the afterglow of GRB~991216 respectively to $\gamma > 150$ and to $\gamma < 150$. For different sources this value may be slightly different. In the E-APE region ($\gamma > 150$) the GRB substructure intensities indeed correlate with the ISM inhomogeneities. In this limited region (see peaks A, B, C) the Lorentz gamma factor of the ABM pulse ranges from $\gamma\sim 304$ to $\gamma\sim 200$. The boundary of the visible region is smaller than the thickness $\Delta R$ of the inhomogeneities (see Fig.~\ref{opening_ETSNCF} and Tab.~\ref{tab1}). Under this condition the adopted spherical approximation is not only mathematically simpler but also fully justified. The angular spreading is not strong enough to wipe out the signal from the inhomogeneity spike.

As we descend in the afterglow ($\gamma < 150$), the Lorentz gamma factor decreases markedly and in the border line case of peak D $\gamma\sim 140$. For the peaks E and F we have $\gamma\sim 50$ and, under these circumstances, the boundary of the visible region becomes much larger than the thickness $\Delta R$ of the inhomogeneities (see Fig.~\ref{opening_ETSNCF} and Tab.~\ref{tab1}). A three dimensional description would be necessary, breaking the spherical symmetry and making the computation more difficult. However we do not need to perform this more complex analysis for peaks E and F: any three dimensional description would {\em a fortiori} augment the smoothing of the observed flux. The spherically symmetric description of the inhomogeneities is already enough to prove the overwhelming effect of the angular spreading \citep{LongArticle2}.

On this general issue of the possible explanation of the observed substructures with the ISM inhomogeneities, there exists in the literature two extreme points of view: the one by Fenimore and collaborators \citep[see e.g.][]{fmn96,fcrsyn99,f99} and Piran and collaborators \citep[see e.g.][]{sp97,p99,p00,p01} on one side and the one by Dermer and collaborators \citep{d98,dbc99,dm99} on the other.

Fenimore and collaborators have emphasized the relevance of a specific signature to be expected in the collision of a relativistic expanding shell with the ISM, what they call a fast rise and exponential decay (FRED) shape. This feature is confirmed by our analysis (see peaks A, B, C in Fig.~\ref{substr_peak}). However they also conclude, sharing the opinion by Piran and collaborators, that the variability observed in GRBs is inconsistent with causally connected variations in a single, symmetric, relativistic shell interacting with the ambient material (``external shocks") \citep{fcrsyn99}. In their opinion the solution of the short time variability has to be envisioned within the protracted activity of an unspecified ``inner engine'' \citep{sp97}; see as well \citet{rm94,pm98,mr00b,mr00,m01}.

On the other hand, Dermer and collaborators, by considering an idealized process occurring at a fixed $\gamma=300$, have reached the opposite conclusions and they purport that GRB light curves are tomographic images of the density distributions of the medium surrounding the sources of GRBs \citep{dm99}.

From our analysis we can conclude that Dermer's conclusions are correct for $\gamma\sim 300$ and do indeed hold for $\gamma > 150$. However, as the gamma factor drops from $\gamma\sim 150$ to $\gamma\sim 1$ (see Fig~\ref{gamma_sub}), the intensity due to the inhomogeneities markedly decreases also due to the angular spreading (events E and F). The initial Lorentz factor of the ABM pulse $\gamma\sim 310$ decreases very rapidly to $\gamma\sim 150$ as soon as a fraction of a typical ISM cloud is engulfed (see Fig.~\ref{gamma_sub} and Tab.~\ref{tab1}). We conclude that the ``tomography" is indeed effective, but uniquely in the first ISM region close to the source and for GRBs with $\gamma > 150$.

One of the most striking feature in our analysis is clearly represented by the fact that the inhomogeneities of a mask of radial dimension of the order of $10^{17}\, {\rm cm}$ give rise to arrival time signals of the order of $20\, {\rm s}$. This outstanding result implies an apparent ``superluminal velocity'' of $\sim 10^5c$ (see Tab.~\ref{tab1}). The ``superluminal velocity'' here considered, first introduced in \citet{lett1}, refers to the motion along the line of sight. This effect is proportional to $\gamma^2$. It is much larger than the one usually considered in the literature, within the context of radio sources and microquasars \citep[see e.g.][]{mr95}, referring to the component of the velocity at right angles to the line of sight \citep[see details in][]{LongArticle2}. This second effect is in fact proportional to $\gamma$ \citep[see][]{r66}. We recall that this ``superluminal velocty'' was the starting point for the enunciation of the RSTT paradigm \citep{lett1}, emphasizing the need of the knowledge of the {\em entire} past worldlines of the source. This need has been further clarified here in the determination of the EQTS surfaces (see Fig.~\ref{opening_ETSNCF}b) which indeed depend on an integral of the Lorentz gamma factor extended over the {\em entire} past worldlines of the source. In turn, therefore, the agreement between the observed structures and the theoretical predicted ones (see Figs.~\ref{grb991216}--\ref{substr_peak}) is also an extremely stringent additional test on the values of the Lorentz gamma factor determined as a function of the radial coordinate within the EMBH theory (see Fig.~\ref{gamma_sub}).

\acknowledgments

We thank M. Rees for pointing out the necessity of presenting our results, R. Giacconi for suggestions on the wording of the manuscript and an anonymous referee for excellent advices.

\clearpage

\clearpage

\begin{deluxetable}{ccccccccc}
\tabletypesize{\footnotesize}
\tablecaption{For each ISM density peak represented in Fig.~\ref{maschera} we give the initial radius $r$, the corresponding comoving time $\tau$, laboratory time $t$, arrival time at the detector $t_a^d$, diameter of the ABM pulse visible area $d_{v}$, Lorentz factor $\gamma$ and observed duration $\Delta t_a^d$ of the afterglow luminosity peaks generated by each density peak. In the last column, the apparent motion in the radial coordinate, evaluated in the arrival time at the detector, leads to an enormous ``superluminal" behavior, up to $9.5\times 10^4\,c$. \label{tab1}}
\tablehead{
\colhead{Peak} & \colhead{$r$ (cm)} & \colhead{$\tau$ (s)} & \colhead{$t$ (s)} & \colhead{$t_a^d$ (s)} & \colhead{$d_v$ (cm)} & \colhead{$\Delta t_a^d$ (s)} & \colhead{$\gamma$} & \colhead{$\begin{array}{c} {\rm ``Superluminal"} \\ v\equiv\frac{r}{t_a^d} \\ \\ \end{array}$}}
\startdata
A & $4.50\times10^{16}$ & $4.88\times10^3$ & $1.50\times10^6$ & $15.8$ & $2.95\times10^{14}$ & $0.400$ & $303.8$ & $9.5\times10^4c$\\
B & $5.20\times10^{16}$ & $5.74\times10^3$ & $1.73\times10^6$ & $19.0$ & $3.89\times10^{14}$ & $0.622$ & $265.4$ & $9.1\times10^4c$\\
C & $5.70\times10^{16}$ & $6.54\times10^3$ & $1.90\times10^6$ & $22.9$ & $5.83\times10^{14}$ & $1.13$  & $200.5$ & $8.3\times10^4c$\\
D & $6.20\times10^{16}$ & $7.64\times10^3$ & $2.07\times10^6$ & $30.1$ & $9.03\times10^{14}$ & $5.16$  & $139.9$ & $6.9\times10^4c$\\
E & $6.50\times10^{16}$ & $9.22\times10^3$ & $2.17\times10^6$ & $55.9$ & $2.27\times10^{15}$ & $10.2$  & $57.23$ & $3.9\times10^4c$\\
F & $6.80\times10^{16}$ & $1.10\times10^4$ & $2.27\times10^6$ & $87.4$ & $2.42\times10^{15}$ & $10.6$  & $56.24$ & $2.6\times10^4c$\\
\enddata
\end{deluxetable}

\end{document}